Effects of post-annealing and cobalt co-doping on superconducting properties of (Ca,Pr)Fe$_2$As$_2$ single crystals

T. Okada, H. Ogino, H. Yakita, A. Yamamoto, K. Kishio, J. Shimoyama

*Department of Applied Chemistry, University of Tokyo, 7-3-1 Hongo, Bunkyo-ku,Tokyo 113-8656, Japan*

ABSTRACT

In order to clarify the origin of anomalous superconductivity in (Ca,*RE*)Fe$_2$As$_2$ system, Pr doped and Pr,Co co-doped CaFe$_2$As$_2$ single crystals were grown by the FeAs flux method. These samples showed two-step superconducting transition with $T_{c1}$ = 25~42 K, and $T_{c2}$ < 16 K, suggesting that (Ca,*RE*)Fe$_2$As$_2$ system has two superconducting components. Post-annealing performed for these crystals in evacuated quartz ampoules at various temperatures revealed that post-annealing at ~400°C increased the *c*-axis length for all samples. This indicates that as-grown crystals have a certain level of strain, which is released by post-annealing at ~400°C. Superconducting properties also changed dramatically by post-annealing. After annealing at 400°C, some of the co-doped samples showed large superconducting volume fraction corresponding to the perfect diamagnetism below $T_{c2}$ and high $J_c$ values of 10$^4$~10$^5$ Acm$^{-2}$ at 2 K in low field, indicating the bulk superconductivity of (Ca,*RE*)Fe$_2$As$_2$ phase occurred below $T_{c2}$. On the contrary, the superconducting volume fraction above $T_{c2}$ was always very small, suggesting that 40 K-class superconductivity observed in this system is originating in the local superconductivity in the crystal.

**1. Introduction**

Various iron-based superconductors consisting of superconducting Fe*Pn*(*Ch*) and blocking layers have been discovered since 2008[1-7]. Among these compounds, '122' family (*AE*Fe$_2$As$_2$ ; *AE* = alkaline earth metals) attracts intense attention even from a practical viewpoint because of its relatively high $T_c$ and low anisotropy. *RE* doped CaFe$_2$As$_2$ (*RE* = La, Ce, Pr and Nd)[8] shows highest $T_c$ in the 122 family, up to 49 K for *RE* = Pr[9]. In addition, La,P co-doped CaFe$_2$As$_2$ single crystal was reported to show high $T_c$ (~45 K) and a large superconducting volume fraction at 5 K[10].

However, the origin of the superconductivity in (Ca,*RE*)Fe$_2$As$_2$ is still controversial due to its anomalous superconducting properties, such as its small superconducting volume fraction above 20 K[8,9,11], larger anisotropy compared with other 122 compounds[12], and two-step superconducting transition[9,12-15]. Furthermore, any superconducting behavior has not been observed in thin films or polycrystalline samples of (Ca,$RE$)Fe$_2$As$_2$. No superconducting single-crystalline (Ca,$RE$)Fe$_2$As$_2$ grown by the Sn flux method is reported, either, although weak superconductivity with $T_c$ exceeding 40 K is reproducibly observed in single crystals grown by the FeAs flux method[8,9,11]. Therefore, the origin of the superconductivity in (Ca,$RE$)Fe$_2$As$_2$ remains unclear and contributions of interface effect accompanied by local strain originating in As vacancy[16,17] or inhomogeneous distribution of dopants[14] are recently proposed.

In order to clarify the origin of superconductivity in (Ca,$RE$)Fe$_2$As$_2$, Pr doped and Pr,Co co-doped CaFe$_2$As$_2$ single crystals were grown by FeAs flux method. Their superconducting properties including the post-annealing effect were studied, because structural, magnetic and superconducting properties of undoped and Co-doped CaFe$_2$As$_2$ single crystals grown by the FeAs flux method can be controlled by post-annealing[18-20]. Structural and superconducting properties of Pr doped and Pr,Co co-doped CaFe$_2$As$_2$ samples were evaluated through X-ray diffraction and magnetic susceptibility measurements.

## 2. Experimental

Single crystals of CaFe$_2$As$_2$ were grown by the FeAs self-flux method with nominal compositions of (Ca$_{1-x}$Pr$_x$)(Fe$_{1-y}$Co$_y$)$_4$As$_4$. According to the nominal Pr and Co compositions $x$ and $y$, samples were labeled as sample 1 ~ sample 8, which are listed in Table 1. In a glove box filled with high purity Ar gas, Ca shots (Alfa Aesar, 99%) were shaved using a file and mixed with FeAs (Furukawa Electric 99.9%), CoAs (Furukawa Electric 99.9%), and Pr (Kojundo Chemical Laboratory, 99.9%) powders. The mixture was pelletized, loaded into an alumina tube and sealed in an evacuated quartz ampoule. The quartz ampoule was heated to 1000°C in 4 hours, kept at 1000°C for 36 hours and slowly cooled down to 950°C in 50 hours, then it was cooled down to room temperature by switching off the furnace. Plate-like single crystals of CaFe$_2$As$_2$ were mechanically separated from the flux. Some of the as-grown crystals were sealed in evacuated quartz ampoules again and annealed at various temperatures ranging from 300 to 600°C for 3~7 days followed by quenching to room temperature. The constituent phases of the samples were analyzed by X-ray diffraction measurement (RIGAKU Ultima-IV) with Cu-$K_\alpha$ radiation generated at 40 kV and 40 mA. Silicon powder was used as an internal standard to determine the $c$-axis length of the CaFe$_2$As$_2$ crystal. Magnetization measurements were performed by a SQUID magnetometer (Quantum Design MPMS-XL5s). $J_c$ was calculated from the width of magnetization hysteresis based on the extended Bean model, $J_c$ [A cm$^{-2}$] = 20 $\Delta M(H)$ / ($a - a^2/3b$), where $\Delta M(H)$ [emu cm$^{-3}$] = $M^+(H) - M^-(H)$ ($M^-(H)$ and $M^+(H)$ are magnetizations measured with field increasing and decreasing, respectively) and $a$ [cm] and $b$ [cm] are lengths of shorter and longer edges of the rectangular crystal.

## 3. Results and discussion

Plate-like single crystals of $(Ca_{1-x}Pr_x)(Fe_{1-y}Co_y)_2As_2$ with flat shiny surface were successfully grown. Their typical dimensions were ~ 3 × 3 × 0.2 (//c) mm$^3$. Fig. 1(a) shows surface XRD patterns, where only sharp 00$l$ peaks are observed. A single diffraction of peak for each $hkl$ index without splitting (See Fig. 1(b) for sample 8) indicates absence of phase separation with different dopant concentrations. In addition, systematic decrease in $c$-axis length with increasing nominal $x$ and $y$ as shown in Fig. 2(a) suggests that doping levels of both Pr and Co are successfully controlled.

For annealed crystals, the value of full width at half maximum (FWHM) of 008 peak seemed to be smaller. In addition, $c$-axes of the samples annealed at 400°C are longer than those of as-grown ones as shown in Fig. 2(a), while samples annealed at ~600°C have comparable or shorter $c$-axes compared with as-grown ones as indicated in Fig. 1(b) and 2(b). This peculiar increase in $c$-axis length after annealing at 400°C has also been reported in undoped[18,20] and Co-doped[19] $CaFe_2As_2$ single crystals grown by the FeAs flux method. In these previous reports, as-grown $CaFe_2As_2$ single crystals grown out of FeAs flux are considered to have some strains, which decrease the lattice volume as external pressure does. Our results indicate that as-grown Pr-doped and Pr,Co co-doped $CaFe_2As_2$ crystals grown by the FeAs flux method also have the strain. This strain is released by annealing at ~400°C, and it reappears by annealing at ~600°C.

Fig. 3(a) shows ZFC magnetization curves for a Pr-doped $CaFe_2As_2$ crystal, sample 2, where magnetic field is applied parallel to the $c$-axis. The as-grown sample showed superconductivity with $T_c$ of ~42 K. However, the superconducting transition was very broad and the diamagnetic signal is not large enough to ensure the bulk superconductivity considering its large demagnetization factor corresponding to the plate-like shape of the crystal. When magnetic field is applied parallel to the $c$-axis of the crystal with dimensions of ~3 × 3 × 0.2 (//c) mm$^3$, the demagnetization factor is roughly estimated to be ~0.90, by approximating the crystal shape by an ellipse, and diamagnetism of $4\pi M / H$ ~ -10 corresponds to full shielding in this case.

After annealing at 400°C superconductivity in sample 2 almost disappeared, while very weak diamagnetism was observed below 8 K. Interestingly, superconductivity of the crystal recovered by annealing at 600°C. This can be attributed to the release and regeneration of strains in the crystal by post-annealing at 400°C and 600°C, respectively. On the other hand, the superconducting volume fraction of a Pr,Co co-doped $CaFe_2As_2$ crystal, sample 8, became larger after annealing at 400°C as shown in Fig. 3(b). For sample 2 and 8, four samples each were measured to confirm reproducibility of the annealing effect, and they showed very similar behaviors. Other Pr,Co co-doped crystals (sample 3~7) also maintained superconductivity even after annealing at 400°C. In these Pr,Co co-doped crystals, $T_c$s were almost unchanged before and after annealing. The superconducting property of $BaFe_2As_2$ under external pressure gives us a clue to understand this behavior. In the cases of $BaFe_2As_2$ crystals doped with K, P or Co, underdoped samples show large enhancement of $T_c$ under external pressure, while optimally-doped or overdoped samples does not show such changes of $T_c$[21-23]. Hence, we consider that sample 2 was in the underdoped state and therefore it showed large change in superconducting properties before and after annealing at 400°C. On the contrary, in Pr,Co co-doped samples, the electron doping level was intrinsically enhanced by Co substitution for the Fe site, thus superconductivity did not disappear after 400°C annealing.

Fig. 4(a) and (b) show zero-field-cooled (ZFC) magnetization curves of sample 3 and 4, respectively, before and after annealing at 400°C. These measurements were performed under 1 Oe applied parallel to the $c$-axis. After the annealing, these samples showed two-step superconducting transition at $T_{c1}$ of ~25 K for sample 3 and ~32 K for sample 4 and $T_{c2}$ of 12 K for both samples. These $T_{c1}$ and $T_{c2}$ were determined by the onset of diamagnetic transition and the kink of the field-cooled (FC) magnetization curve, respectively, as indicated by arrows in the inset of Fig. 4. Temperature dependences of ZFC magnetization of these samples were very small up to ~6 K, and the transitions at $T_{c2}$ were sharp. It should be emphasized that the magnitudes of observed ZFC magnetization below 6 K were close to the expected values for full shielding when we take the shapes of the measured samples into account. For sample 3 with ~2.6 × 1.7 × 0.1(//$c$) mm$^3$ in dimension, the demagnetization factor is roughly estimated to be ~0.93, and $4\pi M / H$ ~ -14 is obtained by assuming the full shielding state. Sample 4 was thicker than the other samples, with dimensions of ~ 2 × 2 × 0.6 mm$^3$. In this case, the demagnetization factor is estimated to be ~0.66 and expected magnetic susceptibility due to perfect diamagnetism is $4\pi M / H$ ~ -2.9. These results strongly suggest the bulk superconductivity of annealed sample 3 and 4 below $T_{c2}$.

The ZFC magnetization curve of the 400°C annealed sample 4 under 1 Oe applied parallel to the $ab$-plane is shown in Fig. 4(c). Superconducting volume fraction at low temperatures is estimated to be approximately 100 %, which also ensures the bulk superconductivity below $T_{c2}$. However, as shown in the enlarged view, diamagnetism derived from the high-$T_c$ component is not observed in this set up. This is probably due to granular and anisotropic distribution of high-$T_c$ component, as suggested by F. Y. Wei *et al.* [16], and larger penetration depth than thickness of high-$T_c$ regions.

Fig. 5 shows magnetization hysteresis loops at 2 K of as-grown sample 2 and 400°C annealed sample 3, 4 and 8. Widths of magnetization hysteresis $\Delta M$ of sample 2 and 8 were quite small and $J_c$ (2 K, ~0 T) of these samples were estimated to be as low as ~ 1 × 10$^2$ A cm$^{-2}$.

This is attributed to weak superconductivity under magnetic field as indicated in the magneto-optical image of (Ca,La)Fe$_2$As$_2$[24]. On the contrary, sample 3 and 4 showed large magnetization hysteresis and their $J_c$ values at 2 K in ~0 T were calculated to be ~$10^5$ A cm$^{-2}$ and ~$10^4$ A cm$^{-2}$, respectively, which were much higher than that of (Ca,La)Fe$_2$As$_2$[24] and comparable to that of Ca(Fe,Co)$_2$As$_2$[25]. For sample 4, magnetization hysteresis loops were measured for another sample in order to confirm reproducibility of the bulk superconductivity, and it also showed high $J_c$, which is shown in Fig. 6, indicating bulk superconductivity below $T_{c2}$. Therefore, we conclude that sample 3 and 4 annealed at 400°C are bulk superconductors below $T_{c2}$.

Here we would like to discuss the two-step superconducting transition observed in the samples. For Pr-doped and Co-free samples, two-step transition at $T_{c1}$ = 42 K and $T_{c2}$ =16 K was observed in sample 2, while sample 1 did not show clear diamagnetism both before and after annealing possibly due to insufficient carrier concentration. In Pr,Co co-doped samples, sample 3~8 exhibited two-step transition as in the case of sample 4, with $T_{c1}$ = 25 ~ 36 K and $T_{c2}$ < 16 K. This two-step transition has also been found in La-[12,13] and Pr-doped[9,14,15] samples by both susceptibility and resistivity measurements. Considering these facts, we believe that two-superconducting components coexist in the (Ca,$RE$)Fe$_2$As$_2$ system.

Our result on sample 3 and 4 indicates that the low-$T_c$ component of this system originates in bulk superconductivity of (Ca,$RE$)Fe$_2$As$_2$. Fig. 6 shows magnetic field dependences of $J_c$ at 2 K for sample 3~6 annealed at 400°C. For sample 4, we have examined for two crystals as described before. Sample 3 showed very high $J_c$ up to ~$10^5$ Acm$^{-2}$, while its $T_{c1}$ was relatively low ~25 K. On the other hand, the estimated $J_c$ values (at 2 K in low field) for sample 5~8 were ~$10^2$ A cm$^{-2}$, which is too small to ensure bulk superconductivity. This fact implies that the bulk superconductivity of low-$T_c$ phase can be observed in a very narrow range of electron doping level, which is determined by the nominal composition as well as the heat treatment condition. In the heavily doped samples compared to sample 4 ($x$ = 0.07, $y$ = 0.02), the superconducting volume fractions at low temperatures are not very small (30 ~ 90 % at 2 K. See Fig. 3(b) for sample 8), whereas their $J_c$ values are quite low. This is probably due to the collapsed tetragonal (CT) phase appeared in the heavily doped compounds and this phase is considered to suppress superconductivity as suggested in Rh-doped CaFe$_2$As$_2$[26]. It should be noted again that any as-grown sample did not show clear bulk superconductivity possibly due to the strain which also degrades superconductivity.

On the other hand, all the samples exhibited very small superconducting volume fraction above $T_{c2}$ even after post-annealing, suggesting that the transition at $T_{c1}$ is due to a local superconductivity in the crystal[12,14,16,17]. It is possible that intergrowth of iron-pnictide superconductor, such as (Ca,$RE$)FeAs$_2$[27,28], is the origin of 40 K-class superconductivity in this system. However, reported $T_c$ on Pr-doped CaFeAs$_2$ is as low as 20 K so far[エラー! ブックマークが定義されていません。], while (Ca,La)FeAs$_2$[エラー! ブックマークが定義されていません。] and (Ca,La)Fe(As,$Pn$)$_2$ ($Pn$ = P, Sb)[29] exhibits 40 K-class superconductivity. Further explorations are needed to reveal the origin of the high-$T_c$ component in Ca-*RE*-Fe-As system.

## 4. Summary

Pr doped and Pr,Co co-doped CaFe$_2$As$_2$ single crystals were successfully grown by FeAs flux method and post-annealing effects on these samples were investigated. Expansion of *c*-axis after annealing at 400°C indicated existence of strain in the as-grown crystals, which might affect superconducting properties in analogy with external pressure. Superconductivity of Pr doped samples disappeared after 400°C annealing, while Pr,Co co-doped samples maintained superconductivity even after 400°C annealing, suggesting that electron doping level was enhanced by Co substitution for the Fe site. It was revealed that (Ca,*RE*)Fe$_2$As$_2$ system has two superconducting components through magnetic susceptibility measurements. Large diamagnetisms indicating the full shielding state were observed below $T_{c2}$ in some of the co-doped samples after the annealing and they exhibited high $J_c$ of $10^4 \sim 10^5$ A cm$^{-2}$ at 2 K in low field. This results indicate that the low-$T_c$ component ($T_c$ < 20 K) is due to bulk superconductivity of (Ca,*RE*)Fe$_2$As$_2$, while bulk superconductivity was confirmed to appear in a limited electron doping level. On the other hand, any evidences of bulk superconductivity have not been found for high-$T_c$ one thus far, supporting the idea of local superconductivity in the bulky crystal.

Acknowledgement

This work was partially supported by SICORP of Japan Science and Technology Agency (JST).

Figure Captions

Fig. 1 (a) Surface XRD patterns of as-grown sample 2 and 8. The inset shows the photograph of grown single crystal. (b) Enlarged view of the XRD patterns around 64° for as-grown and annealed samples.

Fig. 2 (a) $c$-axis lengths of as-grown and 400°C-annealed $(Ca_{1-x}Pr_x)(Fe_{1-y}Co_y)_2As_2$ single crystals. (b) Effect of annealing temperature on $c$-axis lengths of sample 2 ($x = 0.14$, $y = 0$) and 8 ($x = 0.14$, $y = 0.05$).

Fig. 3 Effect of post-annealing on the superconducting properties of (a) sample 2 ($x = 0.14$, $y = 0$) and (b) sample 8 ($x = 0.14$, $y = 0.05$).

Fig. 4 ZFC magnetization curves for sample 3 ($x = 0.05$, $y = 0.02$) (a) and sample 4 ($x = 0.07$, $y = 0.02$) (b) before and after annealing at 400°C under $H // c$ and a ZFC magnetization curve of sample 4 measured under $H // ab$ (c). The insets show the enlarged views of the magnetization curves.

Fig. 5 Magnetic hysteresis loops measured at 2 K under $H // c$ for as-grown sample 2 ($x = 0.14$, $y = 0$) and 400°C annealed crystals of sample 3 ($x = 0.05$, $y = 0.02$), sample 4 ($x = 0.07$, $y = 0.02$) and sample 8 ($x = 0.14$, $y = 0.05$).

Fig. 6 Magnetic field dependences of $J_c$ at 2 K under $H // c$ for sample 3~6 after annealing at 400°C. For sample 4, two crystals indicated as 4a and 4b were measured.

Table 1  Sample labels for $(Ca_{1-x}Pr_x)(Fe_{1-y}Co_y)_2As_2$ single crystals

| sample # | $x$ | $y$ |
|---|---|---|
| 1 | 0.07 | 0 |
| 2 | 0.14 | 0 |
| 3 | 0.05 | 0.02 |
| 4 | 0.07 | 0.02 |
| 5 | 0.10 | 0.02 |
| 6 | 0.14 | 0.02 |
| 7 | 0.07 | 0.05 |
| 8 | 0.14 | 0.05 |

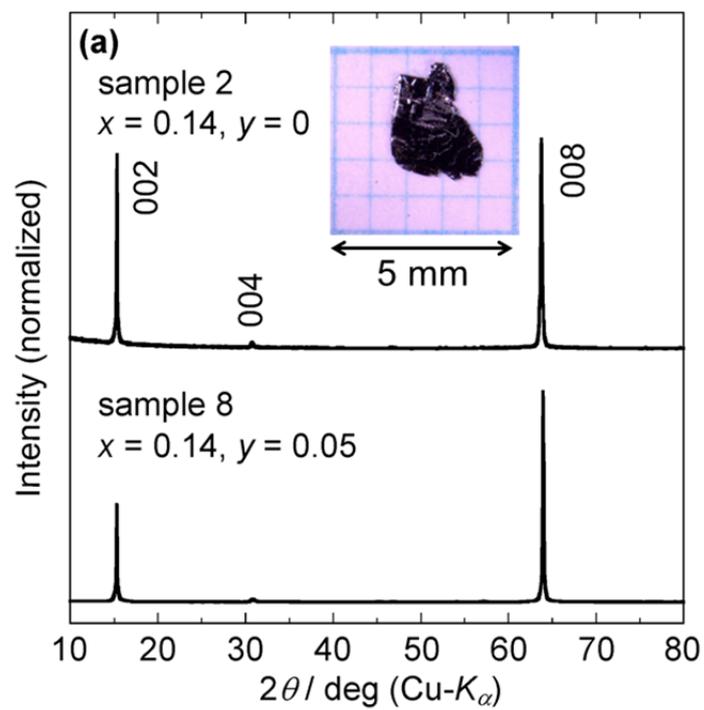

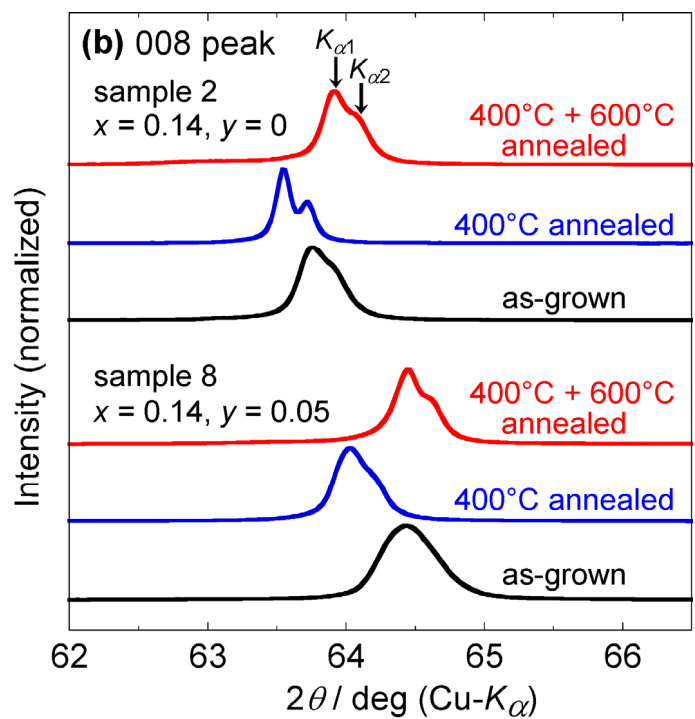

Fig. 1

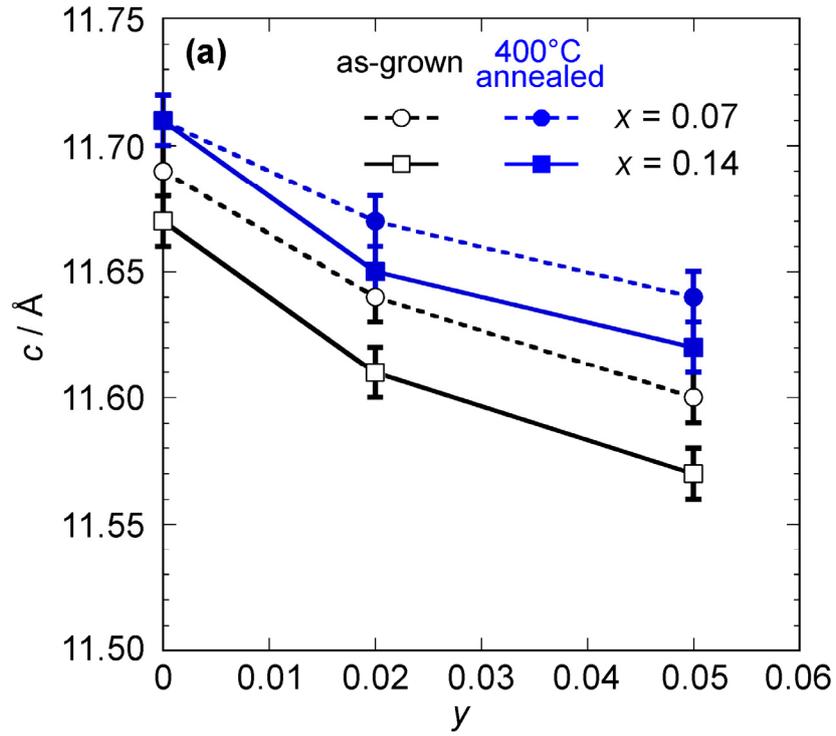

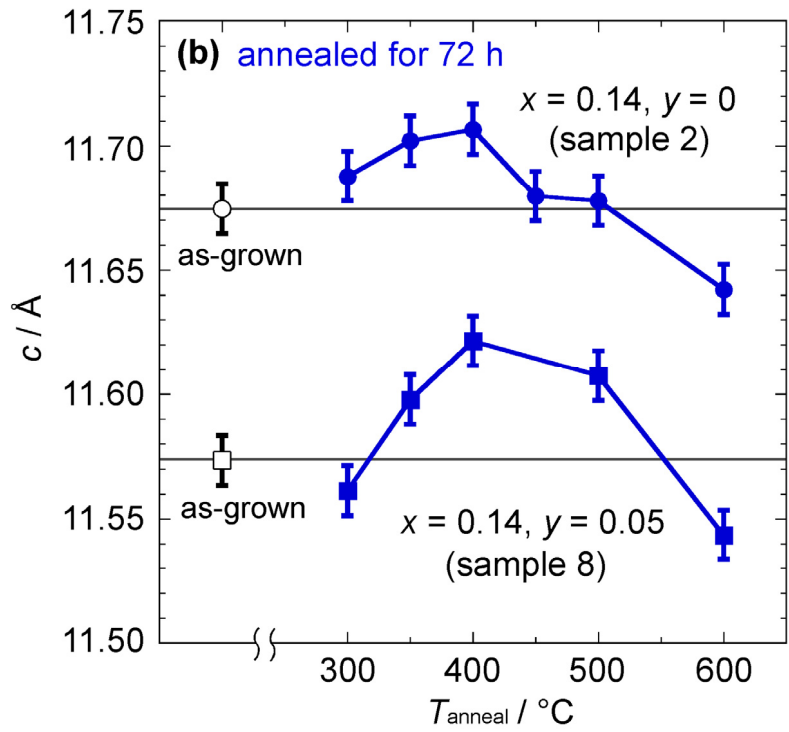

Fig. 2

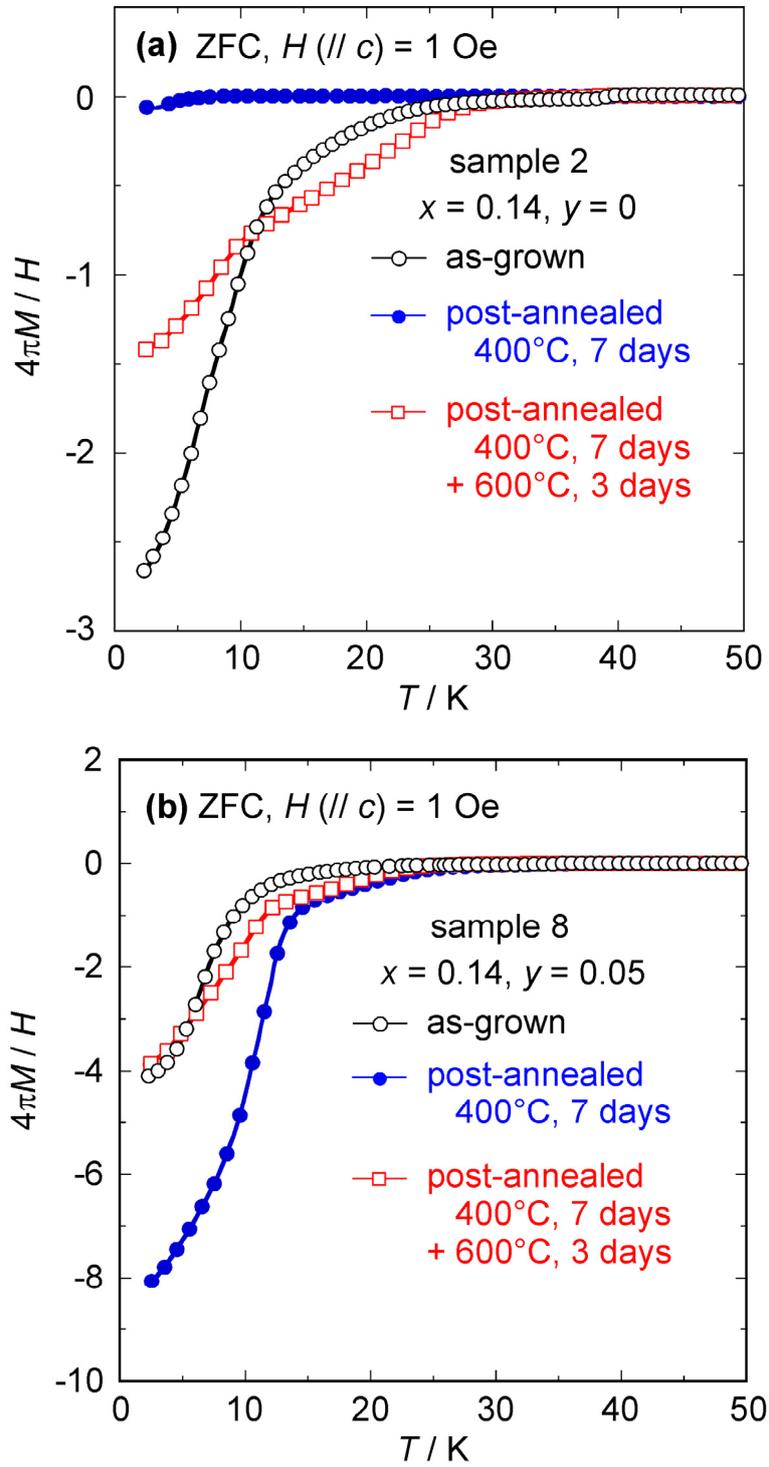

Fig. 3

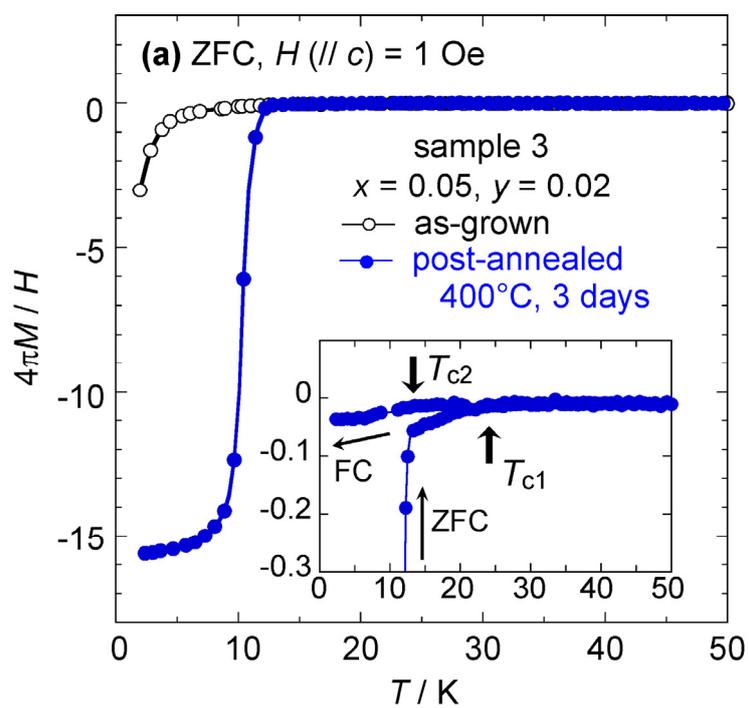

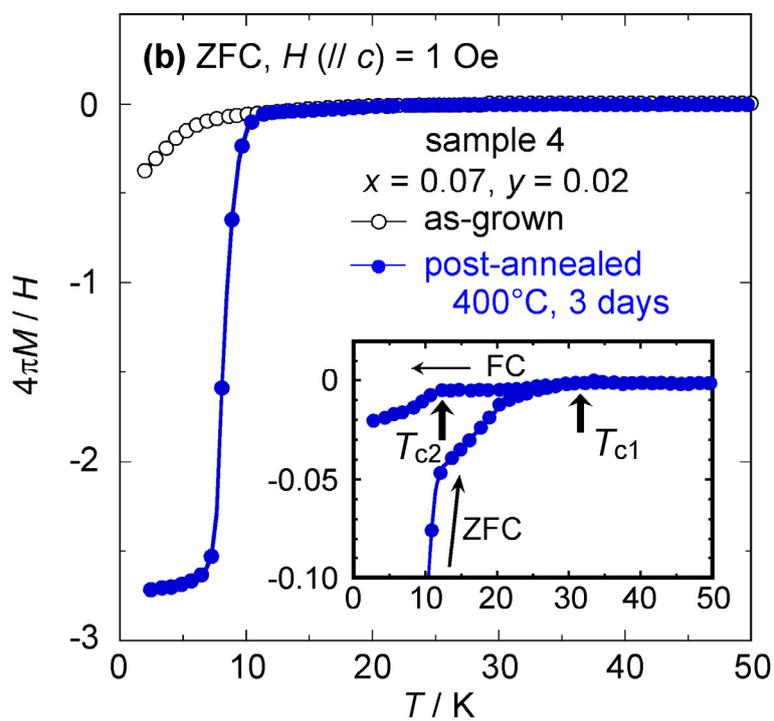

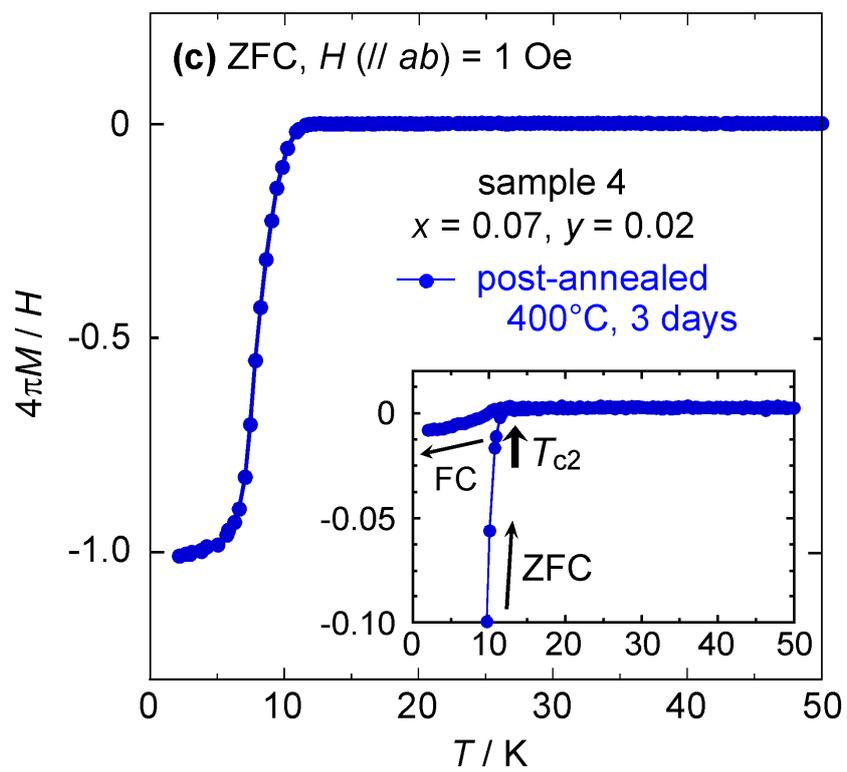

Fig. 4

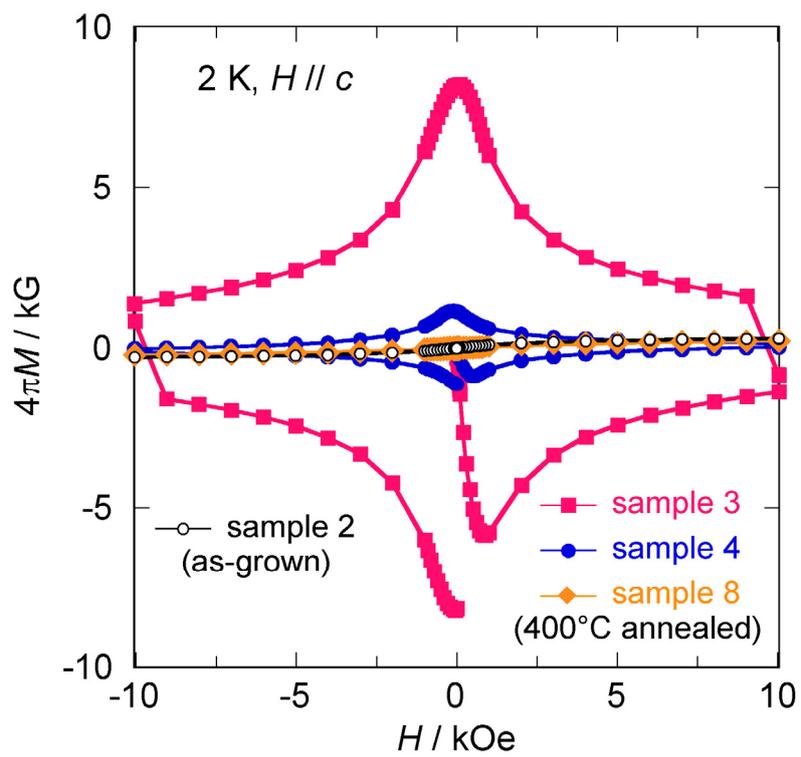

Fig. 5

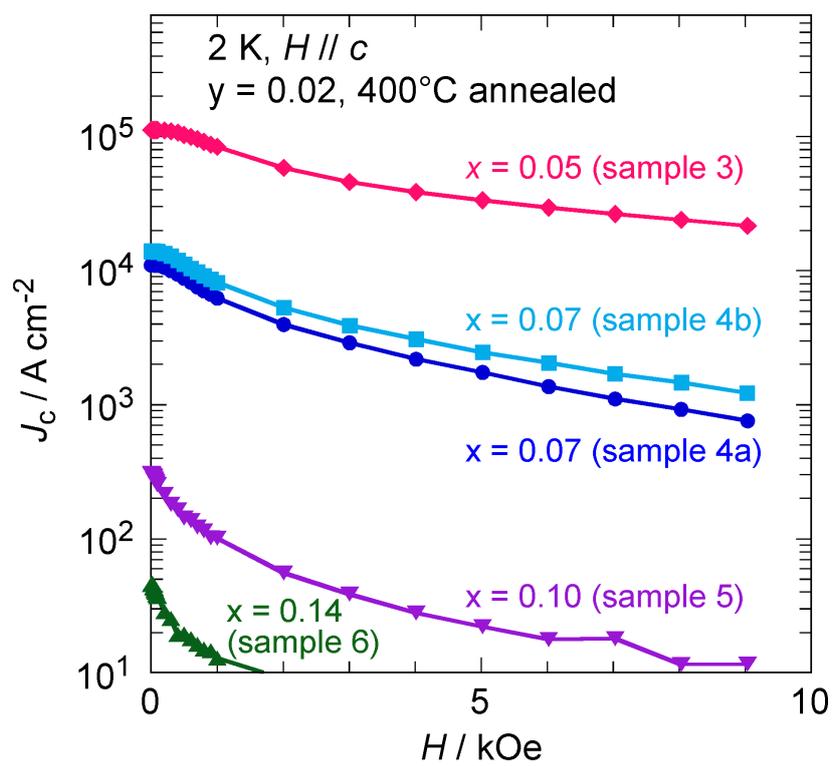

Fig. 6